# Strain Engineering for High-Performance Phase Change Memristors

*Wenhui Hou[1*], Ahmad Azizimanesh[1*], Aditya Dey[2], Yufeng Yang[1], Wuxiucheng Wang[1], Chen Shao[1], Hui Wu[1], Hesam Askari[2], Sobhit Singh[2], and Stephen M. Wu[1,3 †]*

[1]Department of Electrical and Computer Engineering, University of Rochester, Rochester, NY, 14627, USA

[2]Department of Mechanical Engineering, University of Rochester, Rochester, NY, 14627, USA

[3]Department of Physics and Astronomy, University of Rochester, Rochester, NY, 14627, USA

* These authors contributed equally.

† Email: stephen.wu@rochester.edu

**Abstract**

A new mechanism for memristive switching in 2D materials is through electric-field controllable electronic/structural phase transitions, but these devices have not outperformed status quo 2D memristors. Here, we report a high-performance bipolar phase change memristor from strain engineered multilayer 1T'-MoTe$_2$ that now surpasses the performance metrics (on/off ratio, switching voltage, switching speed) of all 2D memristive devices, achieved without forming steps. Using process-induced strain engineering, we directly pattern stressed metallic contacts to induce a semimetallic to semiconducting phase transition in MoTe$_2$ forming a self-aligned vertical transport memristor with semiconducting MoTe$_2$ as the active region. These devices utilize strain to bring them closer to the phase transition boundary and achieve ultra-low ~90 mV switching voltage, ultra-high ~$10^8$ on/off ratio, 5 ns switching, and retention of over $10^5$ s. Engineered tunability of the device switching voltage and on/off ratio is also achieved by varying the single process parameter of contact metal film force (film stress × film thickness).

**Introduction**

Memristors, two terminal resistive switches that can be switched between a high resistance state (HRS) and a low resistance state (LRS) under applied voltage biases[1,2], have been a heavily explored topic of research due to their potential for applications in next-generation memory and neuromorphic computation[3,4]. Memristors based on two dimensional (2D) materials have gained particular interest in recent years due their atomically thin nature, which leads to higher performance metrics such as faster switching, lower power switching, ultra-scalability, and compatibility with flexible substrates[5,6]. While significant advances in memristive devices based on this new materials class have been made[7-19], there still remains a gap between these 2D devices and the highest possible performance of all memristive switching devices from any material. This 2D performance gap results from the typical mechanism of operation, filamentary conduction from the motion of defects, grain boundaries or metal ions, which is the same mechanism as most other memristors. In 2D materials this may lead to many of the same fundamental limitations in device performance such as, needing forming steps, having higher switching voltages or lower on-off ratios, and highly variable device characteristics that cannot be engineered by design.

One way to get around these limiting factors, and to achieve the best overall performance in any memristor, is to change the mechanism of memristive switching beyond filamentary conduction. Recent works have shown that electric field induced phase transitions can occur in 2D $MoTe_2$[12,20], which can switch $MoTe_2$ between the semiconducting (2H) and the semimetallic (1T') phases in a two-terminal memristive device, taking advantage of $MoTe_2$ being the transition metal dichalcogenide (TMD) with the lowest energy difference between different structural/electronic phases among all TMDs[21]. Despite this, the performance of such devices has not exceeded the performance of devices utilizing the status quo mechanisms for memristive switching.

In this article, we demonstrate that using process-induced strain engineering techniques on multilayer MoTe$_2$, we are able to engineer a single-step self-aligned phase-change memristor, with performance metrics showing 90 mV switching voltage, $10^8$ on/off ratio, 5 ns switching, 150 aJ switching energy, and over $10^5$ s retention, in a bipolar switching non-volatile configuration. Our best single device showed combined $7.4 \times 10^7$ on/off ratio at 90 mV switching. This takes advantage of the small energy difference between the two structural states, and uses strain applied to MoTe$_2$ on a device-to-device basis by the contact metal itself to bias the active region closer to the phase-switching point, where a smaller electric field is needed to achieve the same memristive switching. Our devices represent the highest overall performance of any single 2D memristor implementation, achieved without forming steps, with the additional benefit of tunable device parameters (switching voltage, on-off ratio), with the single process variable of contact metal film force (film stress × film thickness). This record high performance represents some of the best overall performance of all memristive technologies, combining ultra-low voltage switching with ultra-high on-off ratios in a fast bipolar (non-volatile) switching configuration with moderately high endurance and retention properties. Our results indicate that further exploration of the phase-change mechanism of memristive switching may lead to one single device implementation offering the best overall performance in any class of memristive device technologies. Successful implementation in this regard could enable robust, ultra-fast, ultra-low-power, and densely scalable applications in non-volatile memories, non-von Neumann/in-memory computing, or implementations of neural-network based neuromorphic computing for artificial intelligence.

Our implementation utilizes process-induced strain engineering techniques, adopted from the standard strain engineering techniques that have been in standard use in commercial CMOS nanofabrication processes. These techniques have been utilized since the 90 nm technology node in 2003 to selectively enhance the mobility of electrons or holes for NMOS or PMOS devices with uniaxial tensile or compressive strain, respectively. Highly stressed thin films, such as $SiN_x$, are selectively deposited on each transistor where the stress of the film relaxes to induce strain in each transistor's channel[22,23]. Our group has pioneered using the same strain engineering techniques on 2D materials based on thin film stress capping layers as a scalable way to apply strain to 2D materials on a device-to-device basis[24-27]. By controlling the film force (film stress × film thickness) and the geometry of the thin film stressor, it has been shown that both the magnitude[26] and the direction[25] of the strain can be controlled. This is done in a defect-free single step deposition process, where we have shown the strain is stable for months to years[27].

In our device design, uniaxial tensile strain is applied by a stressed metallic thin film which doubles as the contact metal for the $MoTe_2$ device (Fig. 1a). Process-induced strain from the metal contacts is transferred from the top down through van der Waals interaction to the underlying 1T'-$MoTe_2$ flake (Fig. 1b), as determined by our previous works on the topic[24-26,28]. Here, this uniaxial strain induces a phase transition underneath the contacts from the semimetallic state to the semiconducting state in a natural vertical transport configuration (Fig. 1c). Thus, with one single process step, we generate a self-aligned phase change memristor structure identical to the original work[12] on the topic (see more details for fabrication in Methods). Since strain is used to induce this phase transition, it is necessarily close to the phase transition point, where small amounts of perturbation from electric-field applied out-of-plane from an applied two-terminal voltage can now induce the phase transition. In fact, this concept was used by our group in the past to generate a three-terminal MoTe$_2$ phase change transistor that operated on gate-controllable strain from a piezoelectric substrate[24]. In this case, process-induced strain from the contacts was used to bias the device out to the most strain sensitive point before piezoelectric strain could be used as the trigger to switch the device between two non-volatile phases. Here, we use the same strain biasing technique, but the trigger is applied electric-field in a two-terminal memristor geometry.

**Strain based resistive switching**

Figure 1d shows the typical nonvolatile bipolar resistive switching I-V characteristic of a strain based 1T'-MoTe$_2$ memristor under 20.7 N m$^{-1}$ film force, where film force (film stress × film thickness) is directly proportional to the amount of strain applied from the metal contacts to the flake, as determined by our previous works on the topic[26,29]. The device starts at a HRS due to the Schottky barrier formed between the metal contact and the phase-switched semiconducting layers. When the voltage between the two metal contacts reaches 110 mV, current suddenly increases and the device enters a LRS. This is because the electric field across the semiconducting layers is large enough to switch them back to the semimetallic state, similar to previous vertical 2H-MoTe$_2$ memristor works[12,20]. Next, by applying a negative voltage of approximately -160 mV between the two metal contacts, the device switches back to the HRS. So, in our devices, the strain-induced phase transition of MoTe$_2$ and the electric-field-induced phase transition of MoTe$_2$ join together in the device switching process, with the former defining the initial semiconducting layers and the latter switching those layers back and forth between the semimetallic and semiconducting phase. This bipolar resistive switching characteristic is repeatable, and no forming process is required. As a control, when devices were made with pure Ag contacts containing low film force (-1.2 N m$^{-1}$), the devices always show low resistance and the I-V curve always shows a linear dependence with high current value, as shown in Supplementary Fig. S1, indicating the MoTe$_2$ flake stayed in the 1T' semimetallic state.

**Strain profile in the device**

To understand how in-plane strain is transferred from the metal contacts to the MoTe$_2$ flake from the interface down through the entire thickness of the flake, we performed Raman measurements on various flakes with different thickness (Fig. 2a), similar to our previous works on process-induced strain on MoS$_2$ and 2H-MoTe$_2$[26,28]. Tensile transparent thin film stressors (Ti/MgF$_2$/Al$_2$O$_3$ or Al$_2$O$_3$/MgF$_2$/Al$_2$O$_3$) with 55 N m$^{-1}$ film force were used to uniformly cover the flakes of different thickness, which not only enables us to apply in-plane biaxial strain to the flakes as shown in our previous works[26,28], but also leaves those same flakes ready to probe through Raman spectroscopic analysis (Fig. 2b). The $A_g^5$ (266 cm$^{-1}$) Raman peak was used to characterize strain, which is a peak shown to be sensitive to strain applied in-plane in the armchair direction of 1T'-MoTe$_2$[30,31]. As shown by the positive $A_g^5$ peak shift in Fig. 2c, we found when uniformly covered by tensile stressed thin films, there is compressive strain in the MoTe$_2$ flake, because the thin film stressor layer relieves its tensile stress by shrinking in all directions, which matches our expectation based on previous works on the topic[26,28]. And more importantly, the largest peak shift was observed on flakes with ~7 nm thickness. Thinner flakes showed smaller peak shift because the bottom layers are clamped by the substrate and cannot be fully strained, an effect that was first shown by our work in MoS$_2$[26]. Peak shift of flakes with thickness larger than 7nm also showed exponential decay because of the finite strain transfer length scale in the out-of-plane direction of 1T'-MoTe$_2$ (~7 nm) and the measured Raman signal is the result of the superimposed Raman signature from highly strained layers of the top 7 nm and the unstrained layers >7 nm away from the top. This finite strain transfer length scale effect is present in all 2D materials, but the length is different between different 2D systems depending on interlayer adhesion[26,28]. It is worth noting from Fig. 2c that points measured from flakes covered by different stressors (Ti/MgF$_2$/Al$_2$O$_3$ or Al$_2$O$_3$/MgF$_2$/Al$_2$O$_3$) can be fit with the same curve, which shows that different stressors with the same film force can apply similar amounts of strain to the covered flake[26].

To understand how strain is transferred from the metal contacts to the MoTe$_2$ flake along the length of the contact, we performed Raman line scans on various flakes with 10 nm thickness and different contact length L, as defined in Fig. 2d. These transparent stressors mimic the geometry of our metallic stressors, allowing us to measure the strain profile applied in this type of contact geometry. Tensile transparent stressors (Ti/MgF$_2$/Al$_2$O$_3$) with 10 N m$^{-1}$ film force were used to partially cover the flake, with the stressor edge perpendicular to the zigzag direction of the flake (Fig. 2e). The $B_g^2$ (105 cm$^{-1}$) Raman peak was used to characterize the strain, which is a peak shown to be sensitive to strain applied in the zig zag direction of 1T'-MoTe$_2$[30,31]. As shown by the negative $B_g^2$ peak shift in Fig. 2f, we found when partially covered by tensile thin films, there is tensile strain in the MoTe$_2$ flake near the stressor's edge, because the edge tends to shrink in the direction perpendicular to itself, which stretches the underlying flake along its zigzag direction, this matches our expectations from our previous works on patterned stressor geometries on MoS$_2$[25]. Compressive strain along the armchair direction of the flake has also been observed, demonstrated by the positive shift of its $A_g^5$ peak (Supplementary Fig. 2), resulting from the positive Poisson ratio of the flake[32]. The relative magnitude of compression along the armchair direction to tension along the zigzag depends on the geometry of our exfoliated flakes, which vary from device to device. A more detailed analysis of these strain profiles is included in the supplementary information. Our comprehensive results show that the edge-induced strain from tensile thin film stressors is directional, which applies tensile strain to the underlying flake in the direction perpendicular to the stressor's edge and applies compressive strain to the flake along the stressor's edge, matching our expectation from previous results on $MoS_2$[25]. And more importantly, when the contact length L increases to 3.2 µm, the magnitude of the $B_g^2$ peak shift and the $A_g^5$ peak shift first increases and reaches a maximum at around 1µm away from the edge, then gradually decreases and drops back to 0 when it is 3.2 µm away from the edge. This shows that the edge induced strain is localized near the edge and decreases quickly when it is away from the edge[25]. Stressors with higher film force (30 N m$^{-1}$) also shows similar in-plane strain transfer profiles (Supplementary Fig. 3). More detailed information about the strain component specific effect on the phase change, as measured through Raman spectroscopy, is included in the supplementary information.

**Effect of process parameters on devices**

With the knowledge of the strain profile in the flake underneath the thin film stressor, we varied different device process parameters and performed a statistical analysis on how these device parameters affect two-terminal electrical device characteristics (Fig. 3).

We first tested the role of contact width W and contact length L (defined in Fig. 3a) by randomly varying W from 0.6 µm to 2.1µm while keeping the film force of the contacts above 15 N m$^{-1}$ for all the devices. We made 258 devices in total and found 55 devices that started at the HRS, which we denote as 'switched devices'. As shown in Fig. 3b, we found the proportion of switched devices showing weak to no dependence on the contact width. Among the 258 devices, there are 150 devices that we also randomly varied in contact length L from 0.6 µm to 2.1 µm during fabrication. As shown in Fig. 3c, among these 150 devices, we found most switched devices having a contact length below 1.1 µm. This is because the contact edge induced strain is localized near the edge, as demonstrated from the Raman profile in Fig. 2f. When contact length is too long, only the region near the edge would switch phase by the applied strain and the device will still be shorted by the unswitched semimetallic 1T' region away from the edge, thereby exhibiting low resistance.

Through the previous analysis, we know to keep contact length <1 µm to design for uniaxial tensile strain. Now fixing the contact geometry, we then tested out the role of angle θ between the edge induced tensile strain and zig zag direction of the flake (defined in Fig. 3d). Due to the in-plane anisotropic nature of the 1T' phase of $MoTe_2$, to change it from the semimetallic phase to the semiconducting 2H phase, it will require tensile strain applied along its zig zag direction, as shown in Fig. 3e, which we modeled using density-functional theory (details can be found in Methods). Of the 258 devices we fabricated above, there are 108 devices that we kept contact length L below 1 µm on purpose during the device fabrication. Within these devices we also randomly varied θ between -90° and 90° relative to the zig zag direction. Among these 108 devices, as shown in Fig. 3f, we found most switched devices having θ between -15° to 15°, which shows that the edge-induced tensile strain must be applied along the zig zag direction of the 1T'-$MoTe_2$ to switch it to the semiconducting state. This characteristic property can be better understood in light of the theoretical findings presented in Fig. 3e in which the structural phase map of the 1T'- and 2H-$MoTe_2$ is drawn. The positioning of the unstrained 1T'-$MoTe_2$ relative to the 2H boundary line requires similar strain directions to cross the phase-change boundary. This strong angular dependence and the match with theoretical calculation strongly suggest that we are correctly inducing a strain-based phase change in our devices that is both applied and detected by the contact metal itself. We are assisted by the fact that our contact metals naturally provide both tensile strain perpendicular to the contact edge (zig zag) as well as compression parallel to the contact edge

(armchair), which according to figure 3e is the fastest way to the phase boundary. More detailed analysis of individual strain component effects on phase change with respect to the calculated phase diagram is presented in the supplementary information.

Finally, understanding the nature of strain applied by the contact geometry, we may now explore the effect of film force, which has been shown in our previous works to be directly proportional to strain magnitude[26,29]. Here, we fixed tensile strain direction to the zig zag direction of the flake, ensured contact length L is below 1.1 µm for uniaxial tension, and varied film force of the contacts by changing contact thickness. New sets of devices with different values of film force from the metal contacts are fabricated, with over 25 devices in each set. Our previous work has shown that by simply changing the thickness of the Ni layers, we can linearly control the film force in Ag/Ni bilayer films[29]. Here, by same method, we show we can also linearly control the film force in Ag/Cr bilayers, as shown in Fig. 3h. Based on this, we fabricated devices made from Ag, Ag/Ni bilayer, or Ag/Cr bilayer films and characterized the percentage of devices starting in the HRS in each set of devices, which we defined as the device yield. From Fig. 3i, it is clear there is a film force and therefore strain magnitude dependence on the nature of the phase-changes that occur under the contacts. Device yield stays 0 % when there is negative film force or low film force applied from the contacts, up to some thresholding value. When film force is larger than 10 N m$^{-1}$, device yield first linearly increases with film force and peaks around 21 N m$^{-1}$. When film force further increases, device yield starts to drop and then increases again, as strain is too large and strain soliton begins to form[33,34]. This is a well-known characteristic of strained 2D materials, where layers begin to slip or ripple on top of each other, before latching on again, producing a sawtooth-like pattern in strain transferred. Looking at other works on the topic[33,34], there is an almost one-to-one relation between our device yield curve, and predicted results of strain transferred given applied stress including soliton formation. This further strengthens the concept that strain from the contacts are generating an active semiconducting region for a self-aligned vertical transport memristor, as summarized in the diagram in Fig. 1a-c.

**Memristive switching characterization**

Next, once all variables are controlled for in the previous section, we explore the current-voltage (IV) characteristics of each of the switched devices. These results are presented in Fig. 4a-d, where each device starting in the HRS exhibits bipolar memristive switching, which occurs without any forming procedure. From the same sets of devices used for film force characterization, we also found the device switching characteristics can be controlled by the film force. Both the on/off ratio and the switching voltage changes with film force applied from the metal contacts. This is understood as higher film force contacts causing higher on/off ratios because more layers and therefore a larger region of the flake were switched to the semiconducting state under the application of larger strain by the contacts. This increasing of the thickness of the active region under the contact makes the HRS of the device less conducting. The highest switching voltage of the device also tends to increase under higher film force, because as more layers of the flake have switched phase by the strain, a larger voltage is required to reach the same electric field needed for the electric-field-induced phase transition. Although we note that the **highest** switching voltage is not representative of the **average** switching voltages observed in our work, which tend to be lower since strain is used to bring the devices closer to the phase transition boundary. For example, for the device presented in figure 4c, switching voltage is 210 mV for a relatively high on/off ratio of $1.4 \times 10^5$. A more detailed summary of the relationship of switching voltage and on-off ratio for all devices fabricated is presented in the next section.

We found that there is statistical variation in each set of the devices, as shown in Supplementary Fig. 4. This is due to uncontrolled device parameters during the fabrication, for example, the shape of exfoliated flakes, which will greatly affect the profile of the strain. Since we are using "as-exfoliated" thin flakes of $MoTe_2$, we can only control contact geometry and therefore there will be variation in device performance due to these still uncontrolled device fabrication variables. When just the highest on/off ratio and the highest switching voltage are extracted from each set of devices, device performance (switching voltage, on-off ratio) clearly scales with the film force we applied in each set (Fig. 5a,b). The highest on-off ratio increases with film force, and eventually reaches over $10^8$, and subsequently decreases because the strain is too large and some regions of the flake are stuck in the semiconducting state and cannot be switched back to the semimetallic state by the electric field. This causes the LRS in these devices to become less conducting. The highest switching voltage increases with film force and eventually plateaus around 1.1 V. By fitting the data from the previous 2H-$MoTe_2$ memristor work[12] (Supplementary Fig. 5), we extracted the relation between the switching voltage and the active region thickness, which shows that 1.1 V switching voltage corresponds to a flake thickness around ~7 nm (Fig. 5b). This matches with the out-of-plane strain transfer length scale in Fig. 2c, showing there is maximally about ~7 nm of the flake that can be switched under high film force. This shows that our expectation of how strain is transferred based on Raman spectroscopic analysis matches device performance, further validating our strain-based phase-change mechanism that drives device design.

We note that while the highest on-off ratio represents the **best** case scenario devices of each set, the highest switching voltage represents the **worst** case scenario devices since it matches the expectation of the original phase change memristor result **without** strain enhancement. The average behavior of our devices have switching voltage far lower than the maximum worst case scenario switching voltage, which is presented as a validity check to the original result. The next section will present in more detail how on-off ratio and switching voltage are not negatively linked, and how further strain engineering may lead to more uniform high performance in devices. Additionally, we also provide a more detailed discussion to exclude other possible switching mechanisms other than the strain-based phase switching in our devices (Supplementary Fig. 6).

**Device performance**

To further understand the performance of these devices, both retention and endurance were tested, where typical results were shown in Fig. 6a,b. LRS and HRS maintained state for longer than $10^5$ s at room temperature and over 270 times DC cycling (Fig. 6a,b). Note that for endurance testing, only IV's that ensured successful switching of the device in every cycle are taken into account[35]. Similarly, devices were tested for switching time where non-volatile device switching was demonstrated with voltage pulses as short as 5 ns with an estimated switching energy, $E_{switch} = I_{LRS} \times V_{swi} \times t_{switc} = 150\ aJ$, presented in supplementary figure 19. To benchmark the overall performance of our devices, comparison of device performance metrics between this work and other recent 2D memristor works[7-19] are given in Fig. 6c,d. All devices fabricated within this work are presented and compared to the highest performing single device of each of the best performing 2D memristor works in recent years in Fig. 6c. As it can be seen, while device variability may still be high due to uncontrolled exfoliated flake geometries, nearly all devices fabricated either match or beat the highest performing status quo 2D memristors known to date. Our devices, with switching voltages as low as 90 mV, On/Off ratios as high as over $10^8$, retention longer than $10^5$ s, endurance over 270 cycles and forming-free switching, shows that within a single implementation we obtain the best overall performance for on/off ratio and switching voltage by several orders of magnitude for all 2D memristive devices without forming. Our best single device is highlighted (red circle) in figure 6c with a combined 90 mV switching voltage and $7.4 \times 10^7$ on-off ratio, where the full IV characteristics presented in supplementary figure 15. Endurance and retention are also moderately high in comparison to other recent works. When compared to the original phase change $MoTe_2$ memristor result (Fig. 6c,d, purple cross), our strain engineering implementation has enhanced all four metrics (on/off ratio, switching voltage, endurance, and retention) by several orders of magnitude, providing strain engineering based enhancement in the same way strain engineering was used to enhance the performance of conventional Si-based MOSFET's. A more detailed comparison for all performance metrics in recent 2D memristors is given in Supplementary Table 1.

Since we have changed the mechanism of memristive switching in our implementation away from filamentary conduction, it also implies that there may also be a change in device area scaling. Incurring larger switching voltages or requiring larger forming voltages as device area becomes smaller is a critical limitation in the commercial implementation of memristive memories[36]. This arises since smaller area devices have less chances to encounter defects to seed filamentary conductivity. Within our own set of fabricated phase change memristors, we see that switching voltage **does not increase** with decreasing device area, hinting at both the power and validity of the phase-change memristor concept (supplementary figure 17). It appears that as device area becomes smaller, switching voltage actually becomes more predictable, with smaller area devices likely having more uniform strain distributions.

Comparing to traditional transition-metal-oxide based memristors, although higher device endurance (>$10^9$) has been demonstrated, they also require higher switching voltage (~1V) and have lower switching speeds because of the filamentary switching mechanism and the thicker active layer to prevent current leakage[37,38]. Our device, with a 2D active layer and a phase-switching mechanism, enables much lower switching voltage and high switching speeds. Endurance may be improved after further device engineering since phase switching is less prone to defect-induced device break down compared to filamentary switching[39]. Similarly, other limiting factors may be further explored through modeling and simulation of strain induced phase changes, such as understanding ideal strain profiles through control of flake-geometries and ideal strain magnitudes to maximize device starting state. Since the mechanism of memristive switching is entirely different, the limitations of previous classes of memristive devices may be overcome, but further work needed to understand the ultimate ceiling of phase-change memristive devices in general.

**Conclusions**

We have reported a strain-based forming-free 1T'-MoTe$_2$ memristor showing excellent overall device performance among 2D memristors. The device combines strain-induced phase transitions and electric-field induced phase transitions of MoTe$_2$ together, with the former setting the initial semiconducting active region and the latter achieving the reversible resistive switching behavior. Both types of phase transition in MoTe$_2$ have been experimentally reported before[12,24], but when the effects are utilized together the benefit is compounded. The contact geometry dependence and the strain direction dependence of the device's starting state match with the strain profile in the flake and the film force tunable device switching characteristics underlines the great engineering potential of this type of strain-based memristor. By utilizing this new type of phase change mechanism, further optimization may provide a realistic path for 2D phase change memristors to achieve the overall best performance of all memristive technologies and provide a realistic pathway to ultra-low-power high-performance applications in memory or neural-network based neuromorphic computing.

**ACKNOWLEDGEMENTS**

We wish to acknowledge support from the National Science Foundation (OMA-1936250 and ECCS-1942815). Raman spectroscopy was performed at the Cornell Center for Materials Research Shared Facilities (CCMR), and CCMR is supported through the NSF MRSEC Program (No. DMR-1719875). We thank Professor Qiang Lin at the University of Rochester in aiding us with the equipment that made it possible to obtain switching speed test results.

**Data Availability**

The data that support the findings of this study are available from the corresponding author upon reasonable request.

**Methods**

**Device Fabrication**

For all devices, MgO substrates with polished surface (Ra<0.5 nm) were oxygen plasma cleaned at 100 W and 200 mTorr for 20 min. The reactive ion chamber was also cleaned by oxygen plasma for 25 min before the substrate was put in to ensure that no contaminants can be redeposited onto the MgO surface[40]. Substrate pre-preparation matters greatly in ensuring high-adhesion of the MoTe$_2$ flake to the substrate, a critical parameter in ensuring predictable strain profiles. Right after the plasma cleaning (<1 min), commercially purchased 1T'-MoTe$_2$ (HQ Graphene) was exfoliated onto the pre-treated MgO substrate using Scotch tape, and the substrate was baked at 100 °C for 90 s with the tape still in contact. The tape was slowly removed after the heating procedure is completed. It should be pointed out that all flakes were exfoliated by the tape from one rectangular-shaped 1T'-MoTe$_2$ crystal, with the long edge of the crystal along its zig zag direction. When the exfoliation was completed, flakes on the substrate showed a preferred cleavage direction along that long edge of the bulk crystal, which is the zigzag direction, confirmed by polarized Raman spectroscopy. We then test substrate adhesion through ultrasonic bath delamination threshold testing[40] by placing the MgO substrate in an acetone and then isopropanol ultrasonic bath for 30 min and 5 min respectively to remove the flakes on the substrate surface with poor adhesion. Optical contrast was then used to quickly identify the rest of the flakes with thickness in the 7nm to 25nm range on the substrate. Ag, Ag/Ni bilayers (25 nm Ag with various thickness Ni on top), Ag/Cr bilayers (25nm Ag with various thickness Cr on top), or Cr/Au bilayers (various thickness Cr with 20nm Au on top) contacts were then patterned on those flakes using direct-write laser photolithography using a Microtech LW405 laserwriter system and S1805 photoresist. All metals are deposited using e-beam evaporation under a chamber pressure below $2\times10^{-5}$ torr at a rate of 1 Å s$^{-1}$. Vacuum condition is maintained throughout the whole deposition process.

**Film force characterization**

Metal thin films are deposited onto cleaned microscope coverslips using the same evaporation conditions as the fabrication of the MoTe₂ memristors. The radius of curvature of the coverslips before and after the deposition are measured using contact profilometry. The average film stress, $\bar{\sigma}$, can then be calculated from the Stoney equation[29]. The applied film force $F_f$, i.e., force per unit width, is obtained through $F_f = t_f \times \bar{\sigma}$, where $t_f$ is the thickness of the metal thin film.

**Device Characterization**

For I-V characteristics, the devices were measured by a Keysight 2901A Precision Source/Measure Unit (SMU) with a current compliance of 1 mA in air. For retention testing, the current was measured every 10 s under 50 mV using the same SMU. For endurance testing, sequences of full I-V sweeps of the device were collected using the same SMU to make sure the device switched in every cycle[35]. The current at 100 mV in every cycle was extracted in Fig. 6b. The retention and endurance test were conducted in a helium gas protected environment to prevent the device from degrading. For switching speed test, Highland technology P400 pulse generator and Cascade Microtech Summit 12000 RF probe station is used. All measurements were performed at room temperature.

**Raman Spectroscopy**

All MoTe₂ samples for Raman spectroscopy were exfoliated on MgO substrates in the same way as memristor devices. For samples of single point Raman measurements, the optically transparent stressor made of 5nm Ti or 10nm Al₂O₃, 100nm MgF₂, and 10nm Al₂O₃ was e-beam evaporated to uniformly cover flakes with different thicknesses. Bottom layer of Ti was deposited at the rate of 0.1 Å s$^{-1}$ and the bottom Al$_2$O$_3$ was deposited at 1 Å s$^{-1}$. Then MgF$_2$ and Al$_2$O$_3$ layer were deposited at the rate of 2 Å s$^{-1}$ and 1 Å s$^{-1}$, respectively. Chamber pressure was kept below 2×10$^{-5}$ torr throughout the deposition. Both films (5nm Ti/100nm MgF$_2$/10nm Al$_2$O$_3$ and 10nm Al$_2$O$_3$/100nm MgF$_2$/10nm Al$_2$O$_3$) were measured to have 55 N m$^{-1}$ film force. Uniformly covered MoTe$_2$ flakes with various thicknesses were then studied with single-point Raman microscopy with a 532nm, 0.5 mW laser. Raman peak shift due to strain was measured by comparing the peak position of covered flakes to the uncovered flakes of the same thickness.

For samples of Raman line-scan measurements, patterned stressors (5nm Ti/20nm MgF$_2$/10nm Al$_2$O$_3$) containing 10 N m$^{-1}$ film force were deposited onto 1T' flakes with 10 nm thickness to partially cover the flakes, so that the stressor edge was perpendicular to the zig zag direction of the flake. Deposition conditions were same as the samples for single point Raman measurements. Raman line scans were performed along the direction that is perpendicular to the stressor's edge. The laser power was kept below 1 mW to prevent laser induced damage to the flakes. Strain profile was also studied in a second set of MoTe$_2$ (8-11 nm thick) samples partially covered by the stressor (5nm Ti/50nm MgF$_2$/10nm Al$_2$O$_3$) with film force of 30 N m$^{-1}$, which showed a similar strain profile to the samples with lower film force. All Raman peaks were fitted by Lorentzian functions.

**DFT calculation of Phase Diagram**

We performed density functional theory (DFT) simulations using the open-source Quantum Espresso package within the generalized gradient approximation (GGA) scheme utilizing projector-augmented waves (PAW) pseudopotentials[41]. The Perdew-Burke-Ernzerhof (PBE) form along with GGA has been used as the exchange-correlation functional[42]. Total energies of the structures were converged by expanding the wavefunctions in a plane wave basis set up to an energy cutoff and charge density of 45 Ry and 400 Ry respectively. We employed 20×10×1 k-point mesh generated by Monkhorst–Pack to sample the reciprocal space. The systems were considered relaxed until all the atomic forces were less than 0.01 eV/ Å. Initially, the ground state configurations of monolayer 1T' and 2H MoTe$_2$ structures were obtained from their DFT optimized rectangular lattice parameters (a = 6.388 Å, b = 3.445 Å for 1T' and a = 6.146 Å, b = 3.549 Å for 2H). A vacuum spacing of 25 Å is considered along the out-of-plane c-axis to avoid spurious interactions between periodic images in that direction.

To capture the phase transition boundary between 1T' and 2H phases as a function of uniaxial strain, their respective *a* and *b* lattice constants were varied followed by calculation of total energies of the strained systems. While relaxing the structures in the presence of strain, the lattice constant along the strained direction is fixed at a given strain value, while that along the transverse direction is fully relaxed along with the inner atomic coordinates. The applied uniaxial strain ($\varepsilon$) is expressed as $\varepsilon = \frac{l - l_0}{l_0}$, where *l* and $l_0$ are the respective lattice constants of strained and pristine MoTe$_2$ structures. On calculating the total energies as a function of lateral lattice constants produces a phase diagram for monolayer 1T'/2H MoTe$_2$. The energy surface intersection contour plot of the two phases with respect to percentage engineering strain in Fig. 3e shows that 1T' to 2H transition occurs at 1.2% strain along *b* lattice for monolayer (zig zag direction), in agreement with earlier work[43].

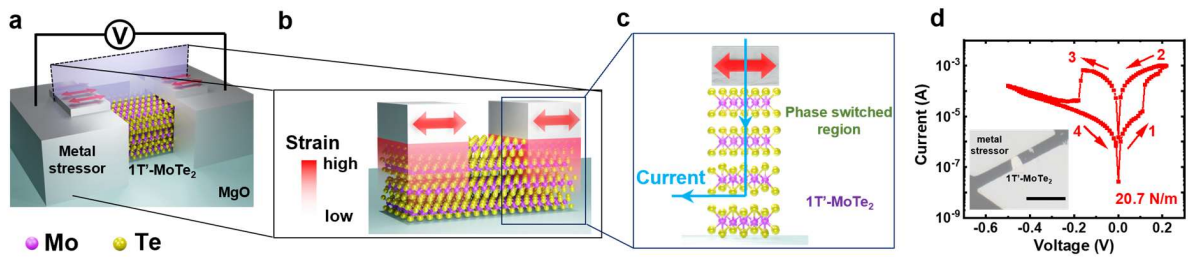

**Fig. 1 Strain-based 1T'- MoTe₂ memristor. a,** Device schematic showing stressed contact metals to MoTe₂, with purple plane representing a device cross-section. **b**, Cross section of the device in **a** with the strain profile. **c**, Mechanism of operation for self-aligned vertical transport based on phase switched MoTe₂ directly beneath the contact metal. Double-headed red arrows in a-c represent the uniaxial edge strain effect. **d**, Typical resistive switching I-V characteristic of the strain-based 1T'-MoTe₂ memristor. Inset: Optical micrograph of the device. Scale bar, 10 μm.

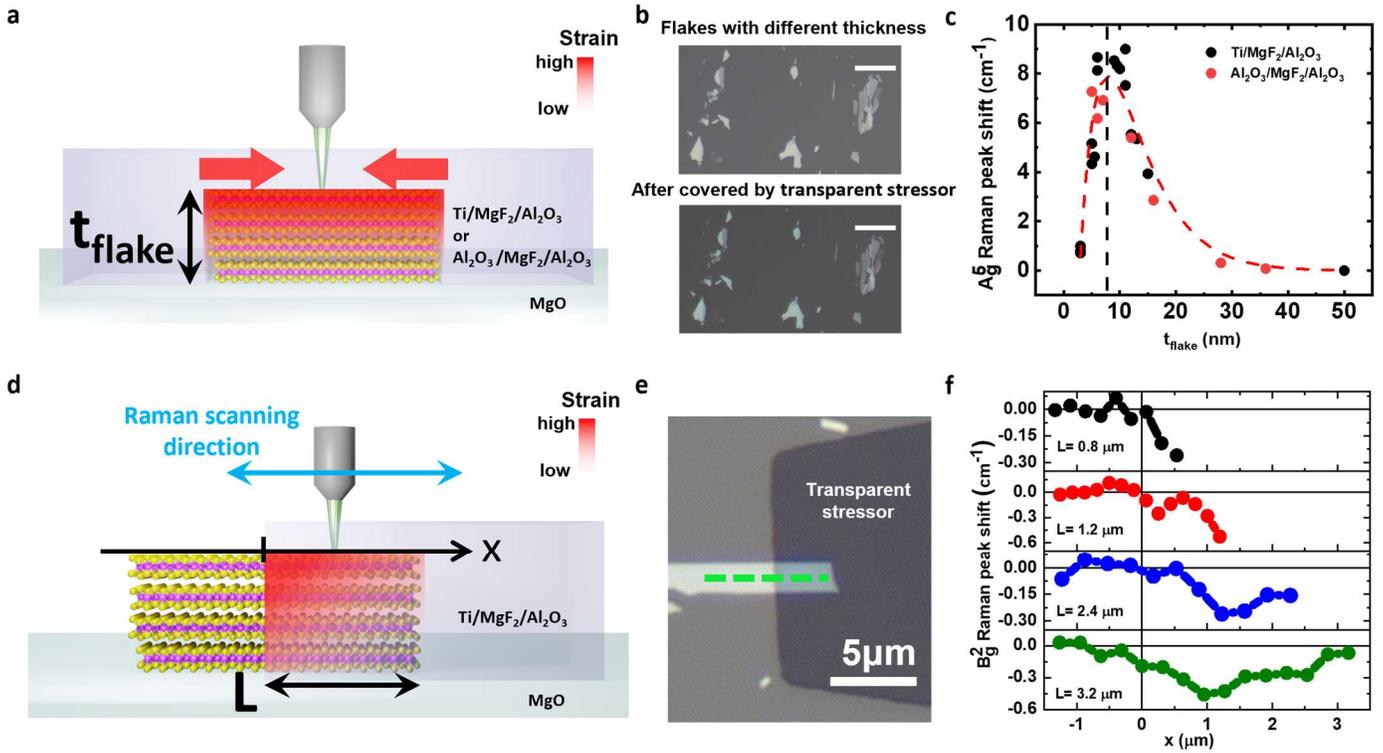

**Fig. 2 Out-of-plane and in-plane strain profile under the stressor. a,** Schematic showing single point Raman measurements setup for out-of-plane strain profile characterization, with variable flake thickness. **b**, Optical microscope image for the MoTe$_2$ flakes before and after encapsulation by the transparent stressor. Scale bar, 10 µm. **c**, $A_g^5$ (266 cm$^{-1}$) Raman peak shift of MoTe$_2$ flakes with different thickness under transparent stressors. **d,** Schematic showing Raman line scan measurements setup for in-plane strain profile characterization, with variable contact length. **e,** Optical microscope image of a MoTe$_2$ flake partially covered by the transparent stressor in the same geometry as contact metals for devices shown in Fig.1d. Green dotted line shows the Raman line-scan path. **f**, $B_g^2$ (105 cm$^{-1}$) Raman peak shift line-scan profile across the stressor edge for flakes with different contact length.

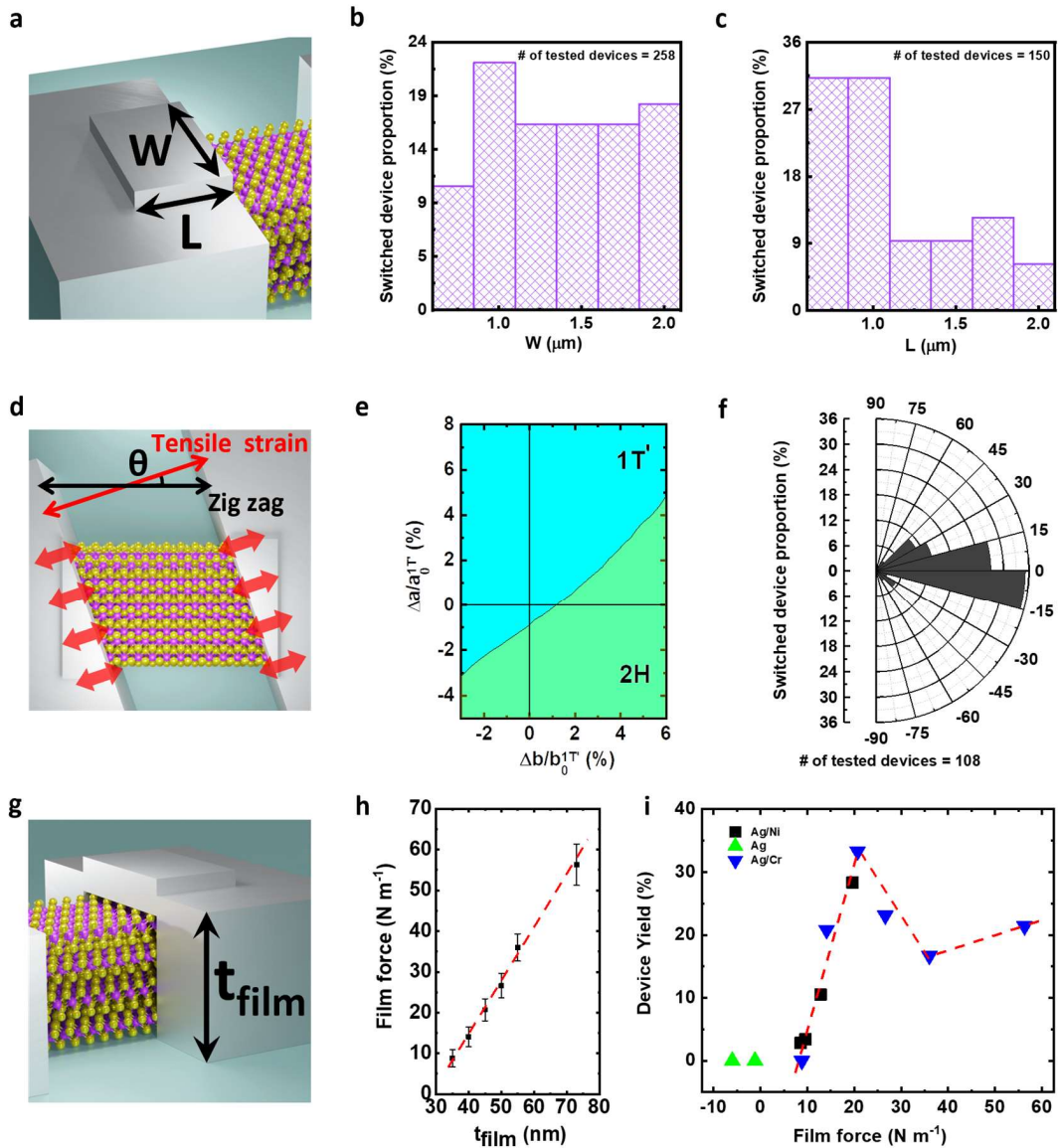

**Fig. 3 Analysis of devices with respect to device parameters. a**, Schematic of the device highlighting variable contact length and contact width. **b,c**, Histogram showing the switched device proportion with different contact width and contact length. **d**, Schematic of the device with the angle of θ between contact edge induced strain and the zigzag direction of the 1T'-MoTe$_2$ flake. **e**, Structural phase diagram of monolayer MoTe$_2$ obtained by DFT, from the phase stability point

of 1T'-MoTe$_2$, where **a** is the lattice constant along armchair direction of 1T' phase and **b** is the lattice constant along zig zag direction of 1T' phase. The origin is placed at unstrained pristine 1T'-MoTe$_2$ in **a** and **b** lattice directions. **f**, Switched device proportion with contact edge induced strain applied to different directions of the 1T'-MoTe$_2$ flake relative to the zigzag direction. **g**, Schematic of the device highlighting the contact thickness. **h**, Film force of Ag/Cr bilayer film as a function of film thickness $t_{film}$, where $t_{film}$ is the total thickness of 25nm Ag and x nm Cr. **i**, Device yield under different film forces where there is a clear film force trend independent of contact metal composition.

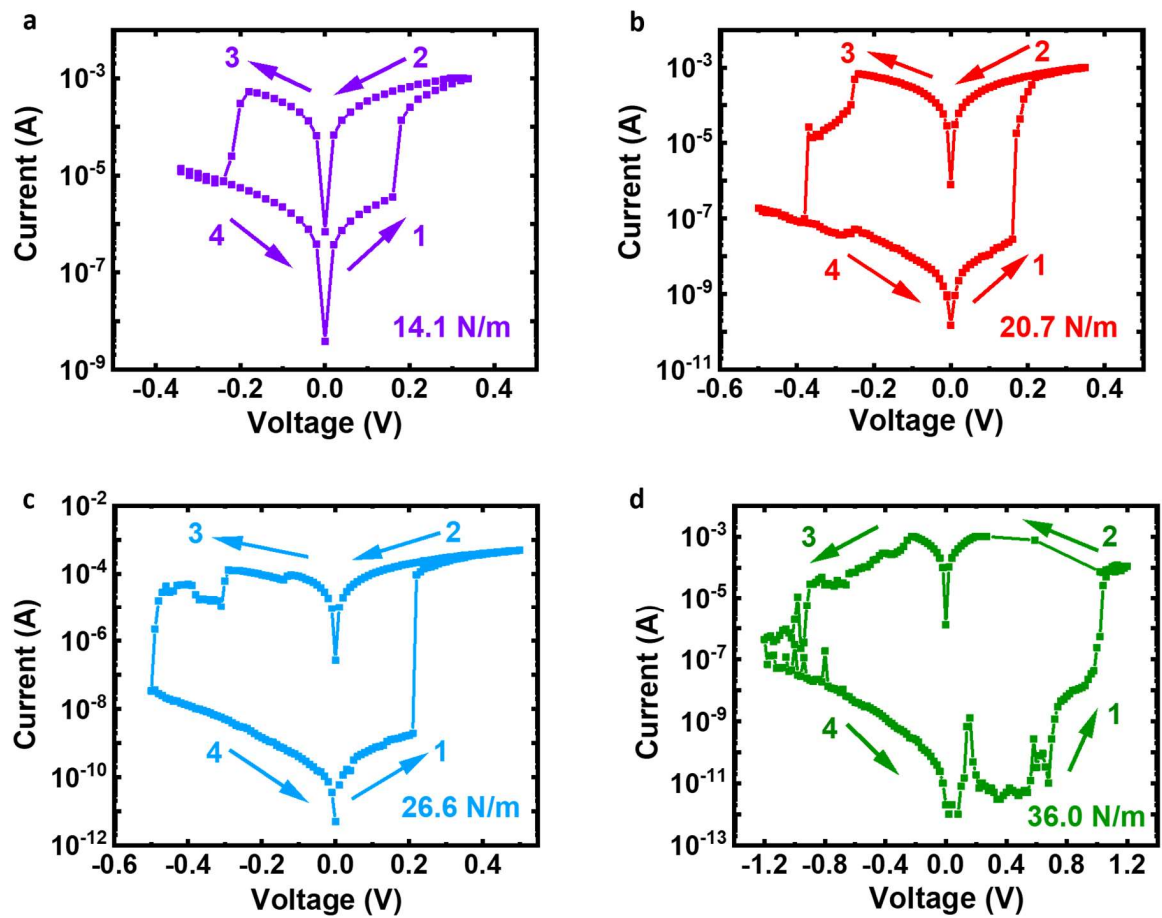

**Fig. 4 Resistive switching under different film force. a-d**, Typical I-V characteristic of devices under 14.1 N m$^{-1}$, 20.7 N m$^{-1}$, 26.6 N m$^{-1}$ and 36.0 N m$^{-1}$ film force, respectively.

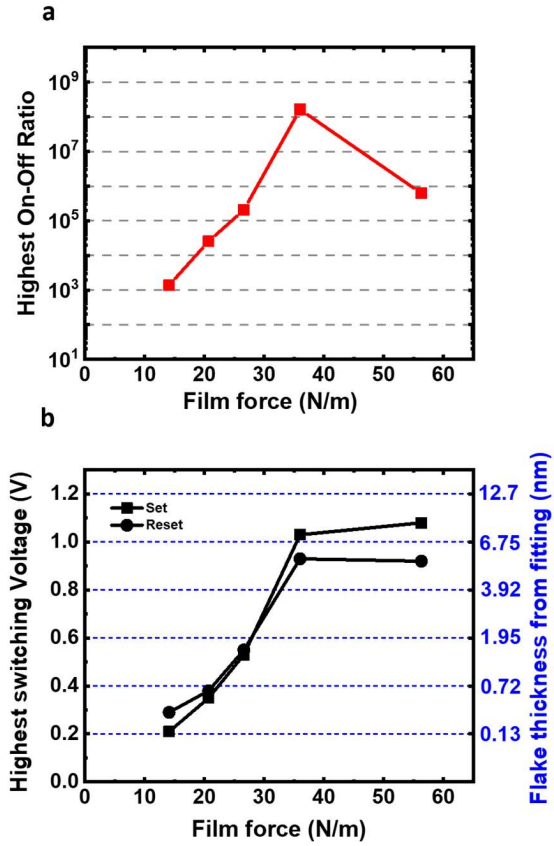

**Fig. 5 Film force controllable switching characteristic.** a, Highest on-off ratio of devices under different film forces. b, Highest switching voltage of devices under different film forces and the corresponding extracted flake thickness assuming thickness scaling from previous MoTe$_2$ phase change works[12], modeled by fitting previous results shown in Supplementary Fig. 5.

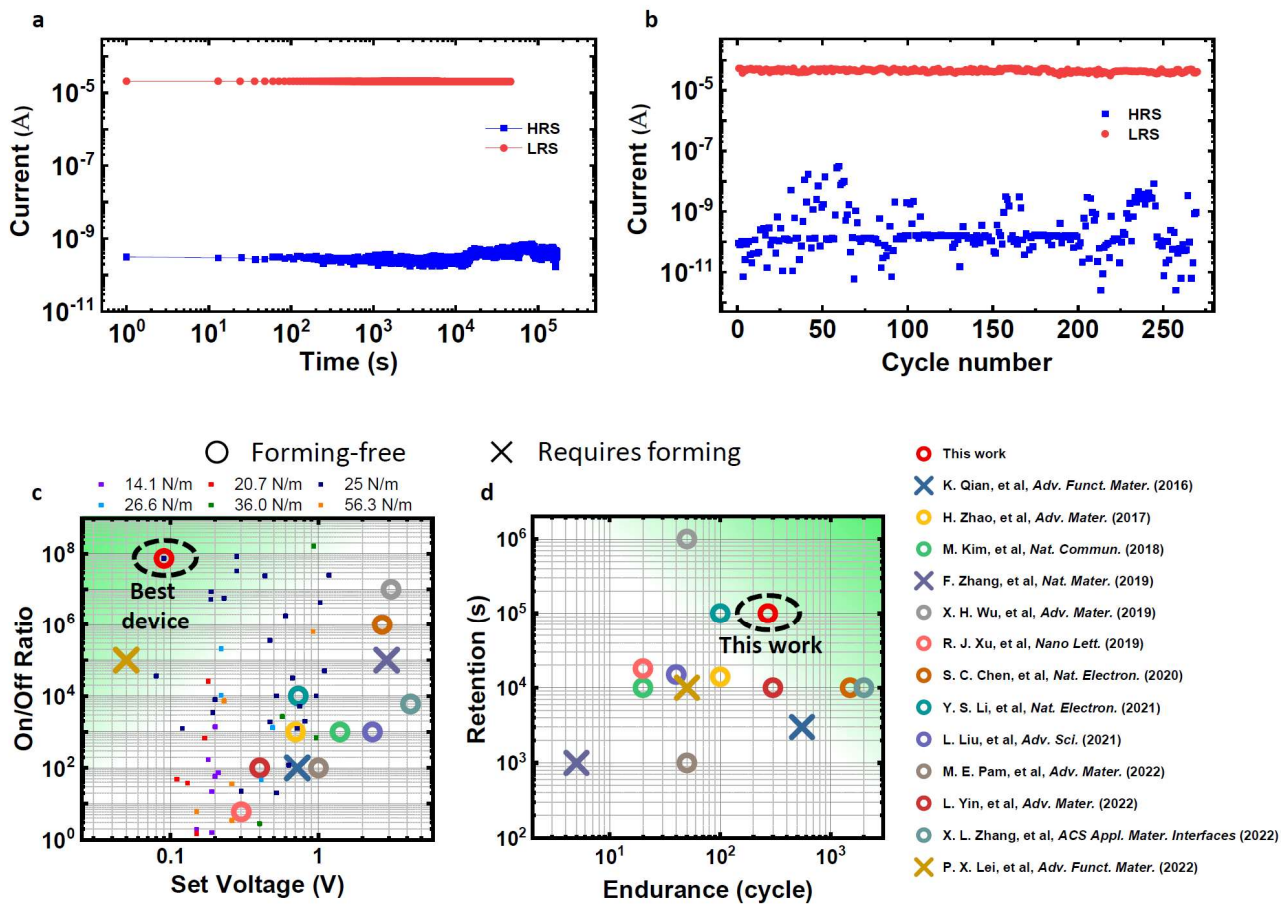

**Fig. 6 Strain-based 1T' MoTe$_2$ memristor performance. a,** Retention of strain based 1T'-MoTe$_2$ memristor over $10^5$s. **b,** Endurance of strain based 1T'-MoTe$_2$ memristor with over 270 manual DC switch cycles. **c,** Comparison of on/off ratio and set voltage of all fabricated devices labeled by film force (filled squares) to highest performance 2D memristors (open circles and crosses). **d,** Comparison of retention and endurance of highest performing 2D memristors. Full comparison table in supplementary information.